\documentclass[aps,prd,twocolumn,superscriptaddress,amsmath,showpacs]{revtex4}
\usepackage[utf8]{inputenc}
\usepackage{booktabs}
\usepackage{longtable}
\usepackage{rotating}
\usepackage{array}
\usepackage{tabularx} 
\usepackage{graphicx}
\usepackage{dcolumn}
\usepackage{bm}
\usepackage{hyperref}
\usepackage{natbib}

\newcommand{\be}{\begin{equation}}
\newcommand{\ee}{\end{equation}}
\newcommand{\bea}{\begin{eqnarray}}
\newcommand{\eea}{\end{eqnarray}}

\newcommand{\WMAP}{{\slshape WMAP~}}

\newcommand{\PLANCK}{{\slshape PLANCK~}}

\newcommand{\Mpc}{{\rm ~Mpc}}

\urldef\mgcamb\url{http://www.sfu.ca/~aha25/MGCAMB.html}
\urldef\planckProducts\url{http://pla.esac.esa.int/pla/aio/planckProducts.html}

\begin{document}
%opening
\title{Updated constraints from the \PLANCK experiment on modified gravity}

\author {Andrea Marchini}
\affiliation{Physics Department and INFN, Universit\`a di Roma ``La Sapienza'', Ple Aldo Moro 2, 00185, Rome, Italy}

\author {Valentina Salvatelli}
\affiliation{Physics Department and INFN, Universit\`a di Roma ``La Sapienza'', Ple Aldo Moro 2, 00185, Rome, Italy}

\begin {abstract}
A modification of the action of the general relativity produces a different pattern for the growth of the cosmic structures below a certain length-scale leaving an imprint on the cosmic microwave background (CMB) anisotropies. We re-examine the upper limits on the length-scale parameter $B_{\rm 0}$ of \textit{f}(R) models using the recent data from the Planck satellite experiment. We also investigate the combined constraints obtained when including the Hubble Space Telescope $H_{\rm 0}$ measurement and the baryon acoustic oscillations measurements from the SDSS, WiggleZ and BOSS surveys.
\end {abstract}

\pacs {04.50.Kd, 04.80.Cc, 98.80.Es, 95.30.Sf}

\maketitle

\section{Introduction} \label{sec:intro}
One of the major challenges for modern cosmology is understanding the 
nature of the cosmic acceleration. Theories that modify general relativity 
in low-density and large-scale regimes are one possible appealing solution 
to the phenomenon, since they can reproduce the accelerated phase in only-matter universes. 

The Cosmic Microwave Background measurements (CMB hereafter) recently provided
from the satellite experiment \PLANCK~\cite{Ade:2013xsa,Planck:2013kta} offer 
a new opportunity to investigate modified gravity scenarios.

It is already well-known that modified gravity (MG) theories deeply influence 
the features of the CMB temperature anisotropies power spectrum (see e.g.~\cite{Acquaviva:2004fv,caldwell2007,daniel2010,calamp}).
As a matter of fact, MG generally introduces modifications at large scales, 
through the late integrated Sachs–Wolfe effect~\cite{giannantonio2009}, and at small scales, 
through the weak lensing effect~\cite{calmod}.

In this brief report we are interested in updating the constraints on MG we have found
from the latest CMB measurements of the South Pole Telescope and the Atacama 
Cosmology Telescope in~\cite{Marchini:2013lpp}, in light of the new high resolution 
CMB data from \PLANCK. For this reason, as in the previous work, we focus on a particular
class of MG models, the \textit{f}(R) theories (see e.g.~\cite{DeFelice:2010aj}), 
exploiting the parametrization proposed in~\cite{MGCAMB2011}. This parametrization fixes 
the background expansion to the standard $\Lambda$CDM scenario and encodes the changes 
in the growth of perturbations in a single parameter $B_{\rm 0}$ that represents 
the length-scale of the theory~\cite{Pogosian:2007sw}.

Other authors have already probed \textit{f}(R) theories trough this kind of parametrization using 
combinations of previous CMB measurements, supernovae luminosity distances, galaxy cluster distribution and 
cluster abundance measurements~\cite{MGCAMB2011,giannantonio2009,Song:2007da,Lombriser:2010mp}. Adding the cluster 
abundance to other data sets actually provides a very tight limit on $B_{\rm 0}$ ($B_{\rm 0}<0.001\; 95$\% c.l.~\cite{Lombriser:2010mp}). 
Our analysis demonstrates that the high resolution CMB measurements from \PLANCK provide strong constraints on 
\textit{f}(R) even without combination with other experiments. Moreover we are interested in evaluating if 
MG alleviates some tensions between the parameter values measured by \PLANCK and other experiments.
 
The report is organized as follows. In Section~\ref{sec:theories} we briefly resume 
the modified gravity model considered in the analysis, in Section~\ref{sec:analysis}
we describe the method of analysis and present the results, in Section~\ref{sec:concl}
we draw our conclusions.

\section {Parametrized Modified Gravity} \label{sec:theories}
In the following analysis we adopt the generic MG parametrization from~\cite{MGCAMB2011} 
where the deviations from the general relativity equations are introduced through two 
parametric functions $\mu(k,a)$ and $\gamma(k,a)$
\begin{equation}
k^2\Psi = - \mu(k,a) 4 \pi G a^2 \lbrace \rho \Delta + 3(\rho + P) \sigma \rbrace \label{mg-poisson}
\end{equation}
\begin{equation}
k^2[\Phi - \gamma(k,a) \Psi] = \mu(k,a)  12 \pi G a^2   (\rho + P) \sigma \label{mg-anisotropy}
\end{equation}
where $\Psi$ and $\Phi$ are the scalar metric potentials in the Newtonian gauge, $\sigma$ is 
the anisotropic stress, $\delta\equiv \delta\rho/\rho$ is the density contrast and $\rho \Delta$
is the comoving density perturbation.
%\begin{equation}
%\rho \Delta = \rho \delta + 3 \frac{Ha}{k}(\rho +P)v \ 
%\end{equation}
%where $v$ is the velocity field.

In~\cite{Bertschinger:2008zb} it has been proved that we can effectively 
reproduce the perturbations of \textit{f}$(R)$ theories choosing the following
parametric form for
$\mu(k,a)$ and $\gamma(k,a)$
\begin{equation}
\mu(k,a)=\frac{1+\frac{4}{3}\lambda_1^2\,k^2a^s}{1+\lambda_1^2\,k^2a^s} \,, \\
\label{BZ}
\gamma(k,a)=\frac{1+\frac{2}{3}\lambda_1^2\,k^2a^s}{1+\frac{4}{3}\lambda_1^2\,k^2a^s} \ 
\end{equation}
where the parameter $s$ must be $\sim 4$ in order to closely mimic the $\Lambda$CDM
expansion~\cite{Zhao:2008bn}.

Therefore, our parametrization has only one degree of freedom that is encoded 
by the length-scale of the theory $\lambda_1$. For scales larger than $\lambda_1$
the dynamic recovers the standard general relativity, otherwise differences in the
potentials $\Phi$ and $\Psi$ are allowed and a different growth pattern for the 
structures can arise. 

In accordance to the previous literature, we present the constraints on the length-scale
in units of the horizon scale, expressing them in terms of the dimensionless parameter $B_{\rm 0}$
\begin{equation}\label{B0}
B_{\rm 0}=\frac{2H_{\rm 0}^2\lambda_1^2}{c^2}
\end{equation}

\section{Analysis and Results} \label{sec:analysis}
We obtain the theoretical CMB power spectrum with the publicly available code 
MGCAMB~\cite{MGCAMB2011} and perform a Markov Chain Monte Carlo analysis with
a modified version of the COSMOMC package~\cite{Lewis:2002ah, Lewis:2013hha}.

Concerning the CMB data set, we consider the \PLANCK measurements\footnote{\textit{http://pla.esac.esa.int/pla/aio/planckProducts.html}}, 
that probes the CMB temperature angular power spectrum up to the multipole 
$\ell=2500$, combined with the CMB polarization measurements performed by the
\WMAP experiment~\cite{Bennett:2012fp} up to the multipole $\ell=23$. We refer
to this combination as PLANCK data set.

We also consider the effect of imposing a gaussian prior on the Hubble parameter 
$H_{\rm 0}=73.8\pm2.4 \,\mathrm{km}\,\mathrm{s}^{-1}\,\mathrm{Mpc}^{-1}$ based on the 
latest Hubble Space Telescope result~\cite{riess2011}. We refer to this prior as HST.

Moreover we take in account a combination of baryon acoustic oscillations (BAO) measurements at different redshifts
from four surveys : the 6dF Galaxy Survey measurement at z = 0.1 provided in~\cite{Beutler:2011hx},
the first SDSS DR7 measurements at $z=0.2$ and $z=0.35$ from~\cite{Percival:2009xn},
the reanalyzed SDSS DR7 measurement at $z=0.35$ from~\cite{Padmanabhan:2012hf}, the
WiggleZ measurements at $z = 0.44, 0.60, 0.73$ extracted in~\cite{Blake:2011en}, the
BOSS DR9 measurement at $z = 0.57$ discussed in~\cite{Anderson:2012sa}. We refer to this
combination as the BAO data set.

The cosmological parameters we sample in the Markov chains are: the MG parameter $B_{\rm 0}$,
the baryon and cold dark matter densities $\Omega_{\rm b} h^2$ and $\Omega_{\rm c} h^2$,
the ratio of the sound horizon to the angular diameter distance at decoupling $\theta$,
the optical depth at the reionization $\tau$, the scalar spectral index $n_{\rm s}$, the
amplitude of the primordial scalar perturbation spectrum $A_{\rm s}$ at $k=0.05\Mpc^{-1}$. 

We also investigate the effect of adding to the former set of parameters the lensing amplitude 
parameter $A_{\rm L}$ that simply rescales the lensing power spectrum 
$C_{\ell}^{\phi \phi} \rightarrow$~$A_{\rm L} C_{\ell}^{\phi \phi}$ as defined in \cite{calabrese2}. 
%\begin{equation}
%C_{\ell}^{\phi \phi} \rightarrow A_{\rm L} C_{\ell}^{\phi \phi}.
%\end{equation}

We fix the helium abundance to $Y_p=0.24$, the number of relativistic degrees of freedom to
$N_{eff}=3.046$, the total neutrino mass to $\sum m_{\nu}=0.06\; eV$. 

In the first part of the analysis we fix the lensing amplitude to $A_{\rm L}=1$ and 
we investigate the results from the PLANCK data set only, PLANCK plus HST prior, 
PLANCK plus BAO measurements. The results we obtain are shown in Tab.~\ref{Tab1}

\begin{figure}[htb!]
\includegraphics[width=4.2 cm,]{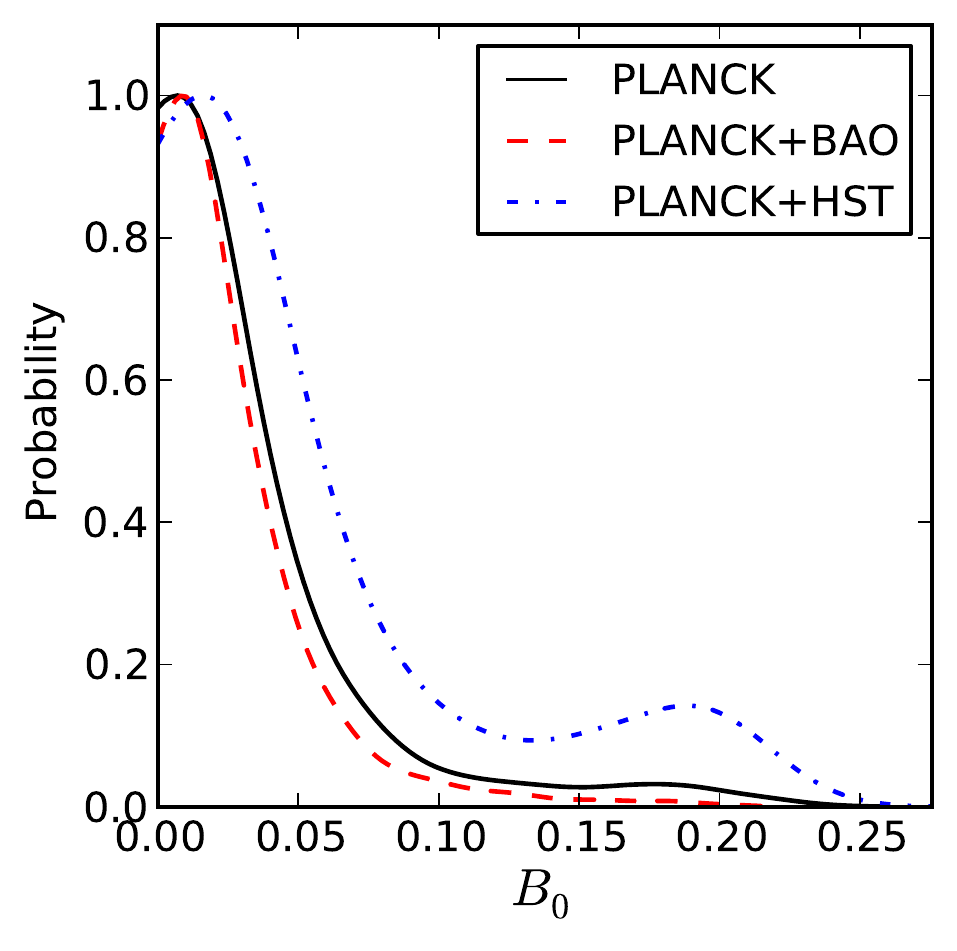}
\includegraphics[width=4.2 cm,]{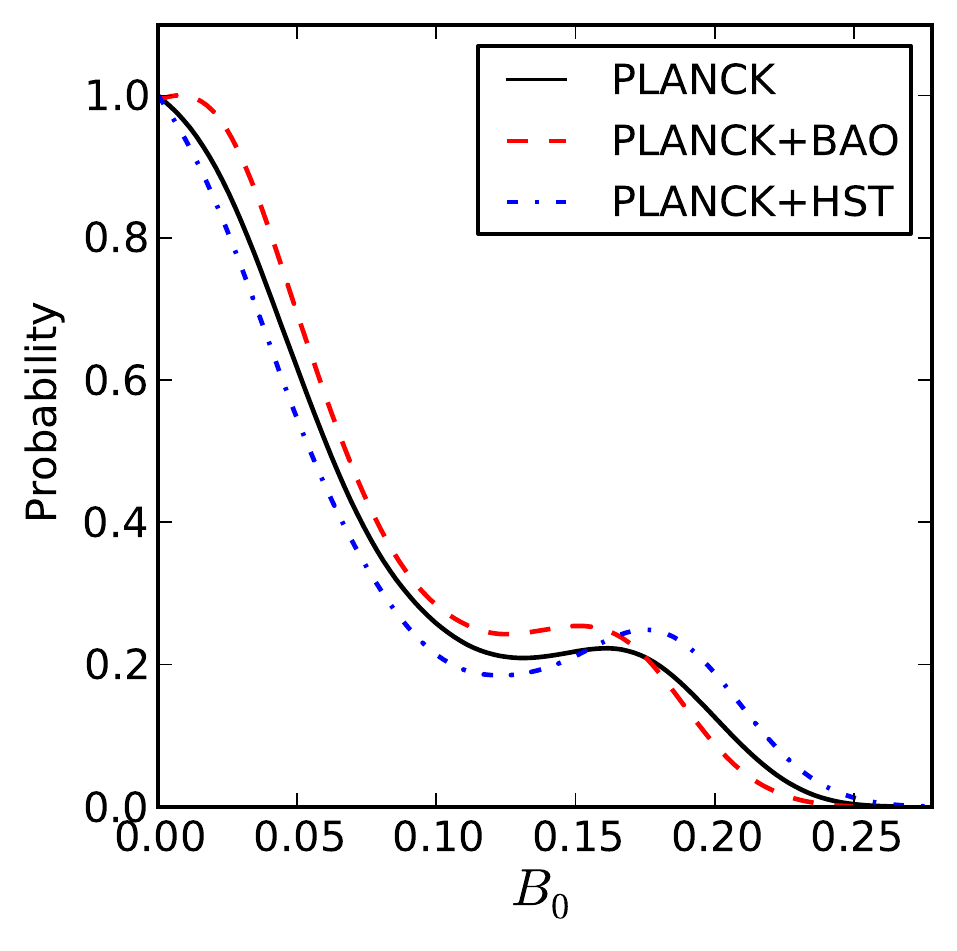}
\caption{\footnotesize Posterior distribution functions for the $B_{\rm 0}$ parameter in the case of $A_{\rm L}=1$ (Left panel) and $A_{\rm L}$ free (Right Panel).
The effect of varying the lensing amplitude is the developing or the increasing of the bimodal behavior for the posterior distribution.}
\label{fig.B0}
\end{figure}

With PLANCK alone we find a constraint on the MG parameter ($B_{\rm 0}<0.134$ at 95\% c.l.)
similar to the constraint obtained in~\cite{Marchini:2013lpp} from the combination
of \WMAP9 and South Pole Telescope measurements. As we expect, the combination PLANCK plus BAO data sets 
provides a tighter constraint ($B_{\rm 0}<0.085$ at 95\% c.l.). As a matter of fact
BAO measurements better constrain the growth of structures at low redshift. The latter combination
also improves the tightest constraint found in \cite{Marchini:2013lpp} by a factor $\sim1.6$.
Instead, when we consider also the HST prior we get a weaker constraint ($B_{\rm 0}<0.195$ at 95\% c.l.). 
This fact is due to the bimodal behavior of the posterior distribution of $B_{\rm 0}$ (see the left panel 
of Fig.~\ref{fig.B0}). 

\begin{figure}[htb!]
\includegraphics[width=4.35cm,]{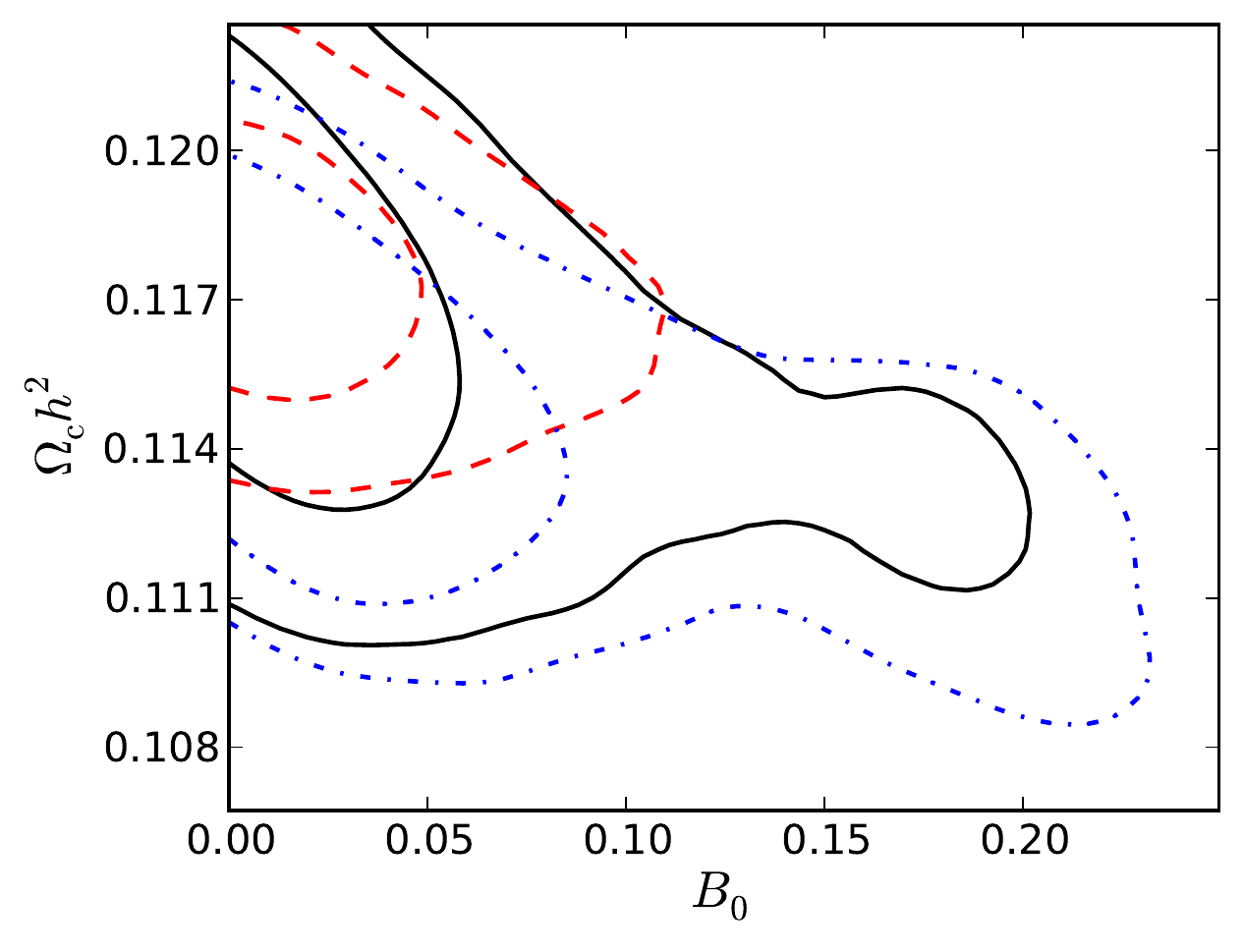}
\includegraphics[width=4.2cm,]{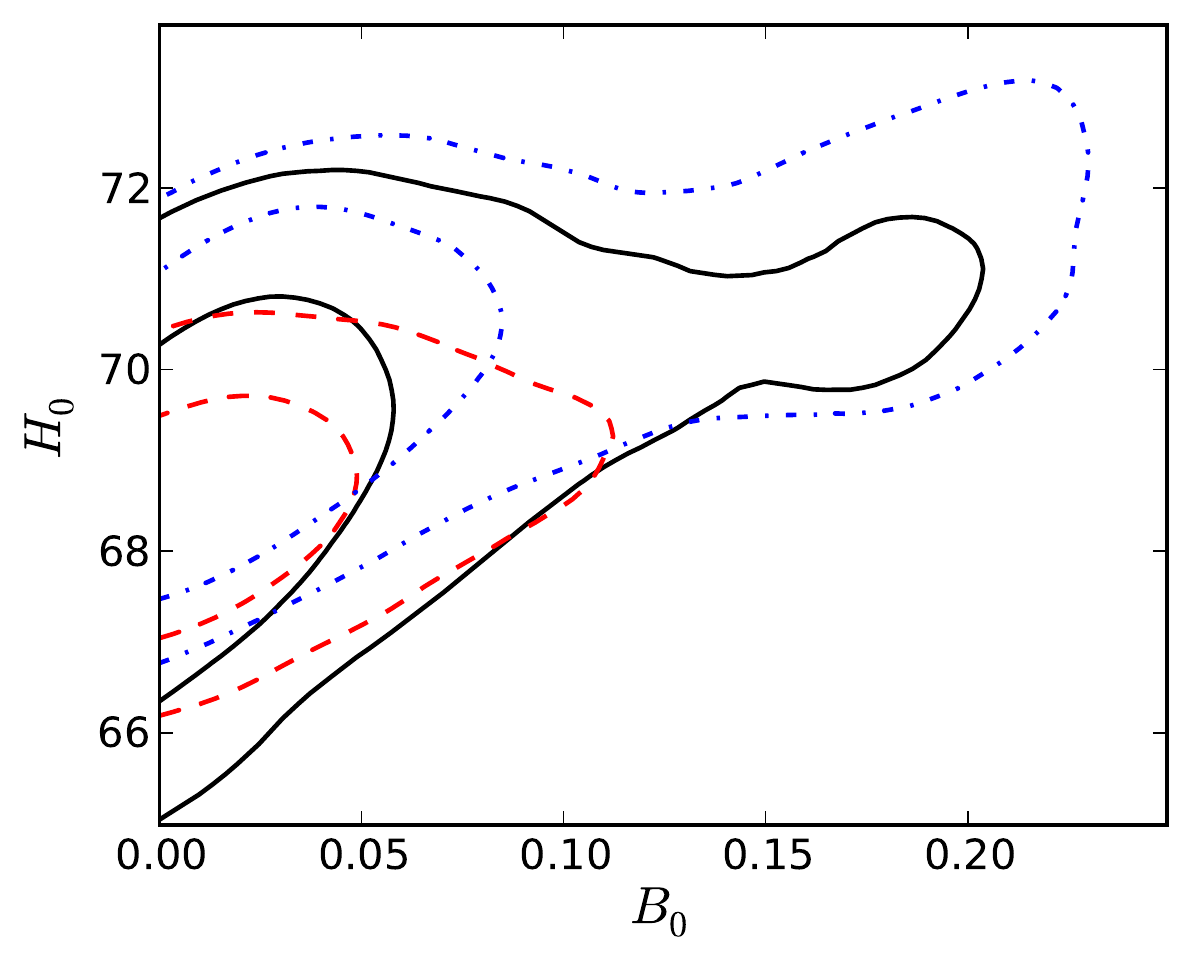}
\caption{\footnotesize 2-D constraints at 68\% c.l. and 95\% c.l. for $B_{\rm 0}$ vs $\Omega_{\rm c} h^2$ and  $H_{\rm 0} $.
We report PLANCK (solid line), PLANCK+BAO (dashed line) and PLANCK+HST (dashed-dotted line). The tension between direct and 
indirect measurements of $H_{\rm 0}$ is clearly alleviated in the MG scenario.}
\label{fig.H0}
\end{figure}

An interesting feature of \textit{f}(R) models is that they seem to alleviate the tension between 
the value of the Hubble constant as inferred by \PLANCK~\cite{Ade:2013lta} ($H_{\rm 0}=67.3\pm1.2\,\mathrm{km}\,\mathrm{s}^{-1}\,\mathrm{Mpc}^{-1}$)
and as measured by HST. As a matter of fact in this scenario the value of $H_{\rm 0}$
increases compared to the $\Lambda$CDM case (see Fig.~\ref{fig.H0} ). The same feature has been already pointed out in quintessence models 
with an interaction in the dark sector (see~\cite{Salvatelli:2013wra} and references therein) that can effectively 
recover scalar-tensor gravitational theories when embedded in the Jordan frame. 

In the second part of the analysis we consider the effects of varying the lensing amplitude $A_{\rm L}$.  The 
results are presented in Tab.~\ref{Tab1}. 
 
When the parameter $A_{\rm L}$ is let free to vary the bimodal behavior of the $B_{\rm 0}$ posterior 
distribution appears also for the PLANCK only and the PLANCK+BAO data set combinations. 
Moreover in the HST case the effect is even increased (Right panel of Fig.~\ref{fig.B0}). 
For this reason the constraints on $B_{\rm 0}$ are weaker respect to the case where $A_{\rm L}$
 is fixed (see Tab.~\ref{Tab1}). 

 The bimodal behavior is due to the fact that the \textit{f}(R) model we consider acts on the CMB power
spectrum in two ways. At high multipoles it imitates the effect of $A_{\rm L}$ greater than one, favouring
lower $A_{\rm L}$ values compared to the $\Lambda$CDM case. At low multipoles instead it lowers the 
integrated Sachs-Wolfe effect plateau, contrary to the effect of an increased $H_{\rm 0}$ value, 
favoring the match between theory and data even in presence of large $H_{\rm 0}$ values. 
The tension between these effects creates the local maximum in the posterior distribution.

Remarkably in MG models the lensing amplitude return to be compatible with $A_{\rm L}=1$
at 68\% c.l. if we consider PLANCK or PLANCK combined with HST and even at 95\% c.l. if we consider
PLANCK combined with BAO. While in the $\Lambda$CDM scenario the standard value $A_{\rm L}=1$ is
compatible with PLANCK and PLANCK combined with BAO only at 95\% c.l. and is excluded at 95\% c.l
when in the combination PLANCK plus HST prior~\cite{Ade:2013lta}.

\begin{table*}[htb!] %\footnotesize
\begin{center}
\begin{tabular}{|l||c c|c c|c c|}
\hline
& \multicolumn{2}{|c|}{{\bf PLANCK}} & \multicolumn{2}{|c|}{{\bf PLANCK+BAO}} & \multicolumn{2}{|c|}{{\bf PLANCK+HST}}\\ \hline
Parameters & Best fit & $68\%$ limit & Best fit & $68\%$ limit & Best fit & $68\%$ limit \\ \hline
$\Omega_{\rm b} h^2$ & 0.02268 & 0.02253 $\pm$ 0.00032 & 0.02265 & 0.02245 $\pm$ 0.00026 & 0.02276 & 0.02274 $\pm$ 0.00030\\
$\Omega_{\rm c} h^2$ & 0.1155 & 0.1165 $\pm$ 0.0027 & 0.1166 & 0.1174 $\pm$ 0.0017 & 0.1141 & 0.1143 $\pm$ 0.0024\\
$100\theta$ & 1.04173 & 1.04189 $\pm$ 0.00066 & 1.04136 & 1.04173 $\pm$ 0.00058 & 1.04202 & 1.04220 $\pm$ 0.00062\\
$\tau$ & 0.081 & 0.087 $\pm$ 0.013 & 0.092 & 0.085 $\pm$ 0.012 & 0.083 & 0.090 $\pm$ 0.013\\
$n_{\rm s}$ & 0.9742 & 0.9697 $\pm$ 0.0076 & 0.9721 & 0.9671 $\pm$ 0.0056 & 0.9767 & 0.9748 $\pm$ 0.0071\\
$\log(10^{10} A_{\rm s})$ & 3.064 & 3.078 $\pm$ 0.025 & 3.089 & 3.077 $\pm$ 0.025 & 3.067 & 3.079 $\pm$ 0.025\\
$B_{\rm 0}$ & 0.024 & $< 0.134$ (95\% c.l.) & 0.022 & $< 0.085$ (95\% c.l.) & 0.036 & $< 0.195$ (95\% c.l.) \\
$\Omega_{\rm m}$ & 0.288 & 0.293 $\pm$ 0.016 & 0.294 & 0.299 $\pm$ 0.010 & 0.279 & 0.280 $\pm$ 0.013\\
$\Omega_{\rm \Lambda}$ & 0.712 & 0.707 $\pm$ 0.016 & 0.706 & 0.701 $\pm$ 0.010 & 0.721 & 0.720 $\pm$ 0.013\\
$z_{re}$ & 10.0 & 10.6 $\pm$ 1.1 & 11.0 & 10.5 $\pm$ 1.1 & 10.2 & 10.8 $\pm$ 1.1\\
$H_{\rm 0} [\mathrm{km}/\mathrm{s}/\mathrm{Mpc}]$ & 69.5 & 69.1 $\pm$ 1.3 & 68.94 & 68.61 $\pm$ 0.79 & 70.2 & 70.2 $\pm$ 1.1\\
$Age/Gyr$ & 13.722 & 13.736 $\pm$ 0.054 & 13.741 & 13.753 $\pm$ 0.039 & 13.698 & 13.696 $\pm$ 0.049\\
\hline
\end{tabular}
%\caption{\footnotesize Best fit values and 68\% c.l. constraints for the \textit{f}(R) models described in Sec.~\ref{sec:theories}
%rom PLANCK data set (first column), from PLANCK+BAO data set (second column) and from PLANCK+HST (third column). 
%nly for the $B_{\rm 0}$ parameter we report an upper limit at 95\% c.l.. The best fit values correspond to the model that
%produces the minimum chi square value, they generally differ from the mean values of parameters at 68\% c.l, unless all
%the posterior distributions are perfectly gaussian.}
%\label{Tab1}
%\label{standard}
%\end{center}
%\end{table*} 
 %\begin{table*}[htb!] \footnotesize
%\begin{center}
\\ 
%\vspace{0.5cm}
\begin{tabular}{|l||c c|c c|c c|}
\hline
 %& \multicolumn{2}{|c|}{{\bf PLANCK}} & \multicolumn{2}{|c|}{{\bf PLANCK+BAO}} & \multicolumn{2}{|c|}{{\bf PLANCK+HST}}\\ \hline
 &  &  & & & &\\ \hline
Parameters & Best fit & $68\%$ limit & Best fit & $68\%$ limit & Best fit & $68\%$ limit \\ \hline
$\Omega_{\rm b} h^2$ & 0.02231 & 0.02241 $\pm$ 0.00035 & 0.02244 & 0.02234 $\pm$ 0.00029 & 0.02274 & 0.02265 $\pm$ 0.00033\\
$\Omega_{\rm c} h^2$ & 0.1180 & 0.1172 $\pm$ 0.0030 & 0.1186 & 0.1180 $\pm$ 0.0017 & 0.1130 & 0.1147 $\pm$ 0.0026\\
$100\theta$ & 1.04169 & 1.04172 $\pm$ 0.00069 & 1.04166 & 1.04159 $\pm$ 0.00057 & 1.04215 & 1.04215 $\pm$ 0.00065\\
$\tau$ & 0.094 & 0.088 $\pm$ 0.012 & 0.084 & 0.088 $\pm$ 0.012 & 0.086 & 0.091 $\pm$ 0.013\\
$n_{\rm s}$ & 0.9689 & 0.9675 $\pm$ 0.0086 & 0.9648 & 0.9655 $\pm$ 0.0060 & 0.9777 & 0.9740 $\pm$ 0.0078\\
$\log(10^{10} A_{\rm s})$ & 3.097 & 3.082 $\pm$ 0.026 & 3.079 & 3.082 $\pm$ 0.024 & 3.068 & 3.082 $\pm$ 0.026\\
$B_{\rm 0}$ & 0.040 & $< 0.185$ (95\% c.l.) & 0.012 & $< 0.175$ (95\% c.l.) & 0.012 & $< 0.198$ (95\% c.l.) \\
$A_{\rm L}$ & 0.90 & $0.91^{+0.10}_{-0.14}$ & 0.981 & $0.89^{+0.092}_{-0.11}$ & 1.04 & $0.96^{+0.10}_{-0.14}$\\
$\Omega_{\rm m}$ & 0.303 & 0.298 $\pm$ 0.018 & 0.305 & 0.303 $\pm$ 0.011 & 0.273 & 0.283 $\pm$ 0.015\\
$\Omega_{\rm \Lambda}$ & 0.697 & 0.702 $\pm$ 0.018 & 0.695 & 0.697 $\pm$ 0.011 & 0.727 & 0.717 $\pm$ 0.015\\
$z_{re}$ & 11.4 & 10.8 $\pm$ 1.1 & 10.5 & 10.8 $\pm$ 1.1 & 10.4 & 10.9 $\pm$ 1.1\\
$H_{\rm 0} [\mathrm{km}/\mathrm{s}/\mathrm{Mpc}]$ & 68.2 & 68.7 $\pm$ 1.4 & 68.16 & 68.28 $\pm$ 0.85 & 70.6 & 69.9 $\pm$ 1.3\\
$Age/Gyr$ & 13.771 & 13.757 $\pm$ 0.060 & 13.761 & 13.771 $\pm$ 0.043 & 13.690 & 13.708 $\pm$ 0.055\\
\hline
\end{tabular}
\caption{\footnotesize Best fit values and 68\% c.l. constraints for the \textit{f}(R) models described in Sec.~\ref{sec:theories}
in the case we fix $A_{\rm L}=1$ (upper table) or we let $A_{\rm L}$ free to vary (bottom table) from PLANCK data set (first column), 
from PLANCK+BAO data set (second column) and from PLANCK+HST (third column). Only for the $B_{\rm 0}$ parameter we report an upper 
limit at 95\% c.l.. The best fit values correspond to the model that produces the minimum chi square value, they generally differ 
from the mean values of parameters at 68\% c.l, unless all the posterior distributions are perfectly gaussian. The constraints on
$B_{\rm 0}$ are weaker in the bottom case because a bimodal behavior appears in the likelihood distributions (see Fig.~\ref{fig.B0})}.
\label{Tab1}
%\label{standard}
\end{center}
\end{table*} 

%\begin{figure}[htb!]
%\includegraphics[width=4.2 cm,]{fr_B0.pdf}
%\includegraphics[width=4.2 cm,]{fr_Alens_B0.pdf}
%\caption{\footnotesize Posterior distribution functions for the $B_{\rm 0}$ parameter in the case of $A_{\rm L}=1$ (Left panel) and $A_{\rm L}$ free (Right Panel).
%The effect of varying the lensing amplitude is the developing or the increasing of the bimodal behavior for the posterior distribution.}
%\label{fig.B0}
%\end{figure} 

%\begin{figure}[htb!]
%\includegraphics[width=4.35cm,]{B0_och2.pdf}
%\includegraphics[width=4.0cm,]{B0_theta.pdf}
%\includegraphics[width=4.2cm,]{B0_H0.pdf}
%\caption{\footnotesize 2-D constraints at 68\% c.l. and 95\% c.l. for $B_{\rm 0}$ vs $\Omega_{\rm c} h^2$ and  $H_{\rm 0} $.
%We report PLANCK (solid line), PLANCK+BAO (dashed line) and PLANCK+HST (dashed-dotted line). The tension between direct and 
%indirect measurements of $H_{\rm 0}$ is clearly alleviated in the MG scenario.}
%\label{fig.H0}
%\end{figure} 

\begin{figure*}[htb!]
\includegraphics[width=3.0cm,]{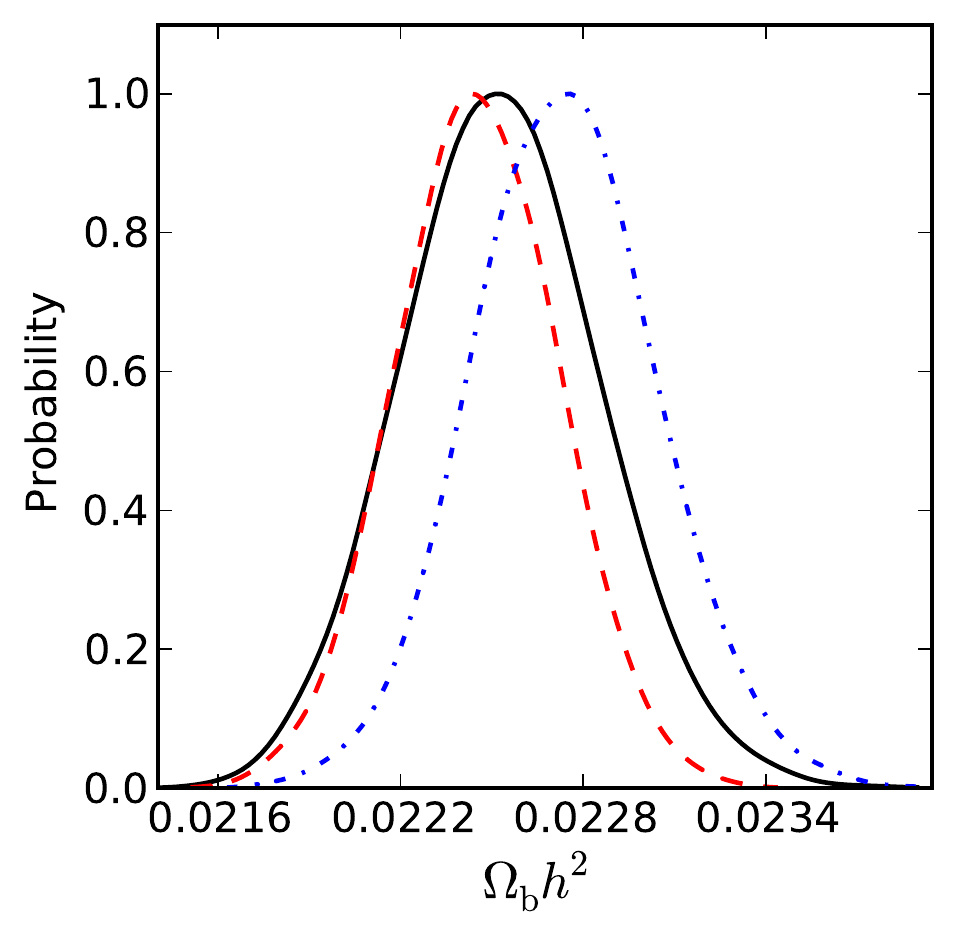}
\includegraphics[width=3.0cm,]{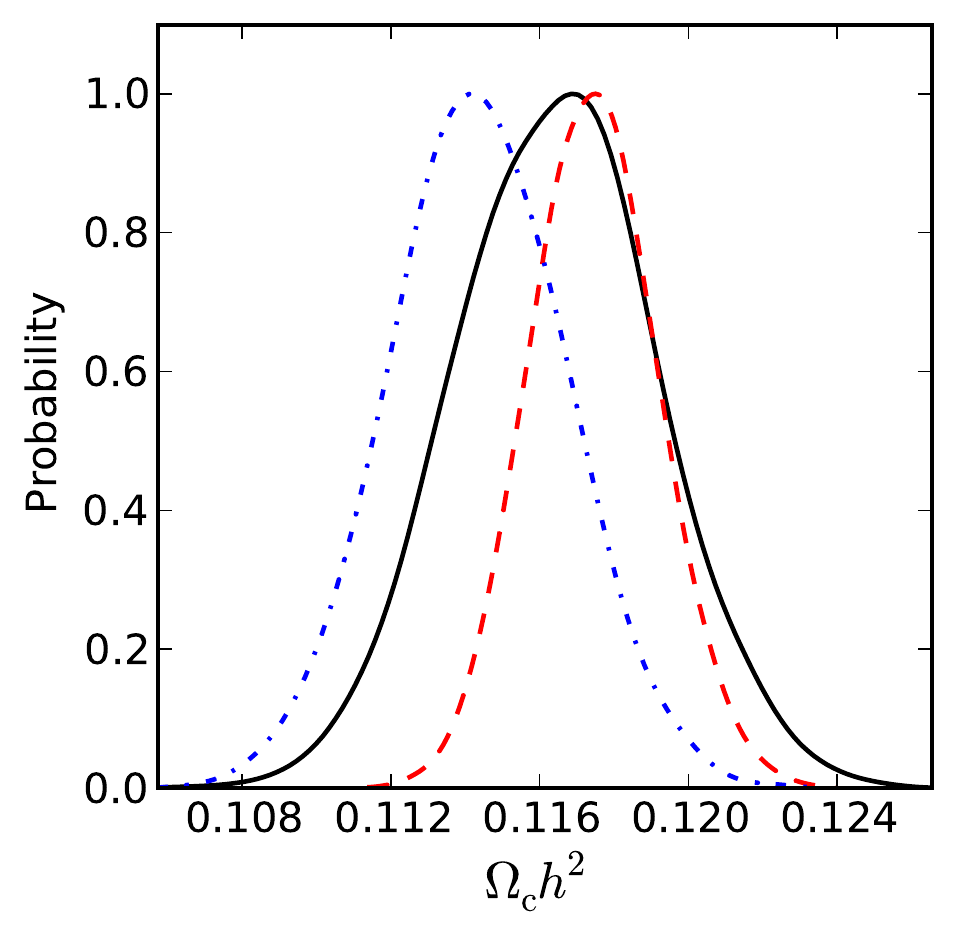}
\includegraphics[width=3.0cm,]{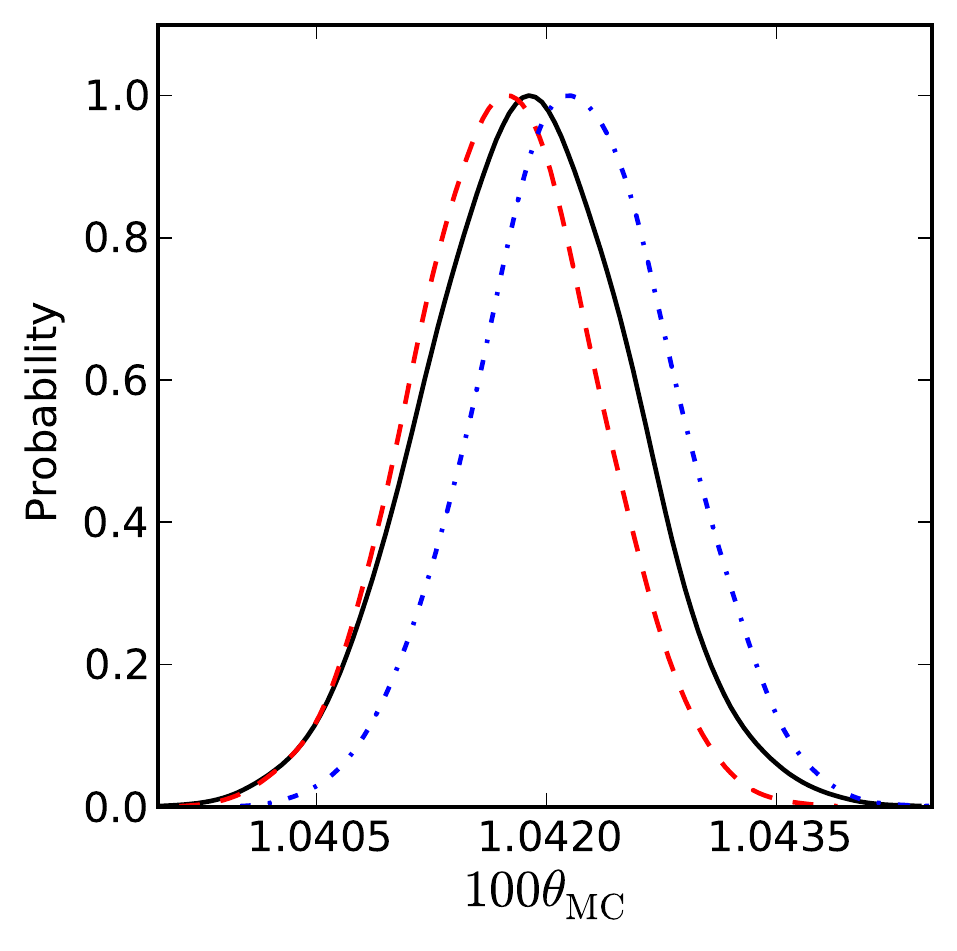}
%\end{figure*} 
%\begin{figure*}[htb!]
\includegraphics[width=3.0cm,]{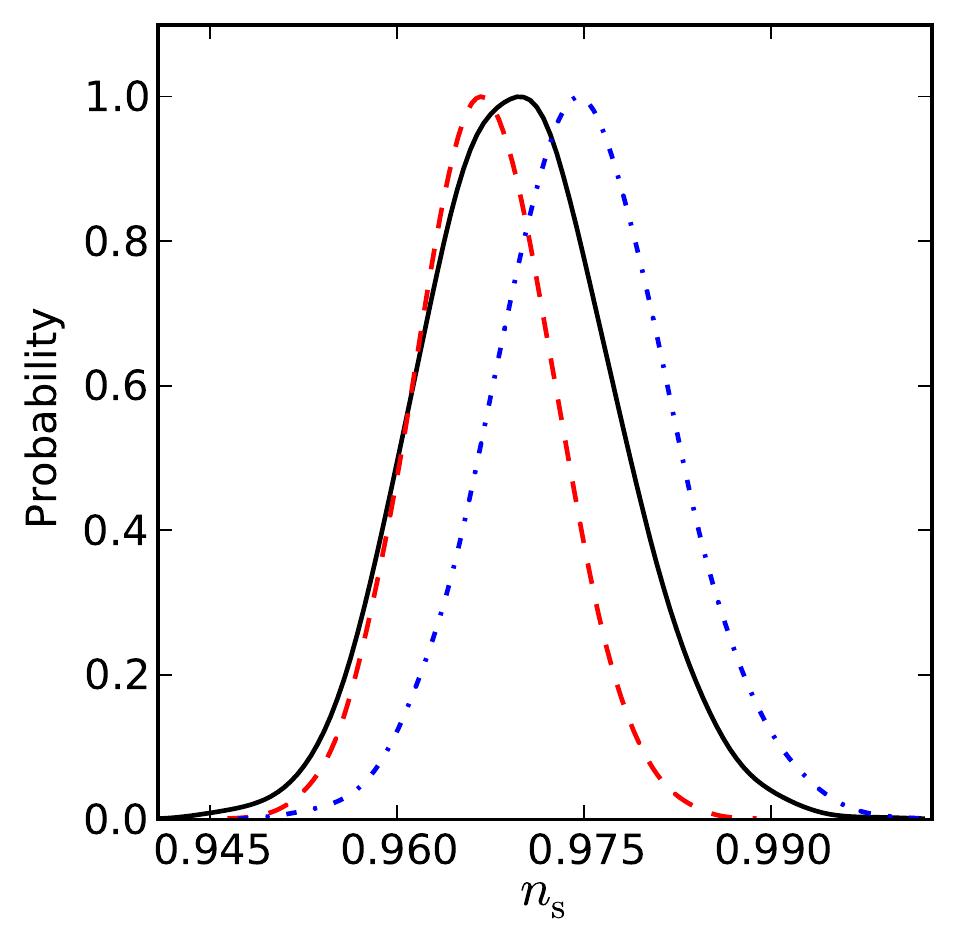}
\includegraphics[width=3.0cm,]{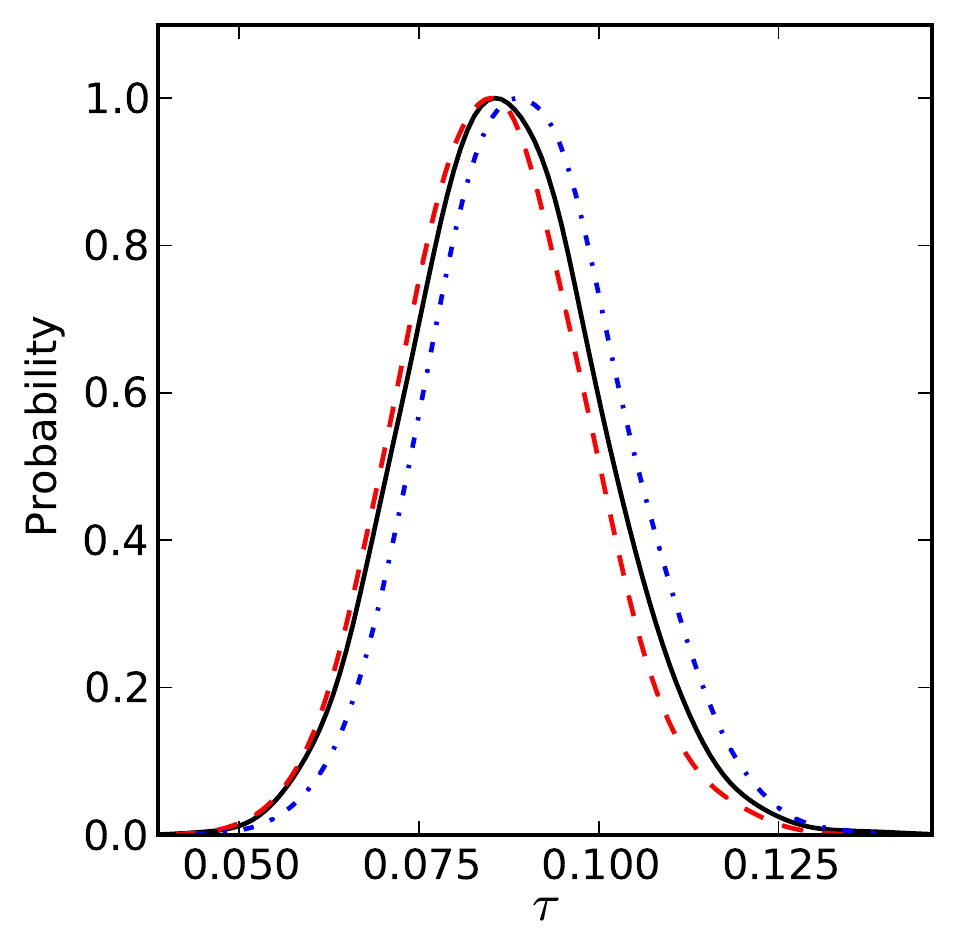}
\includegraphics[width=3.0cm,]{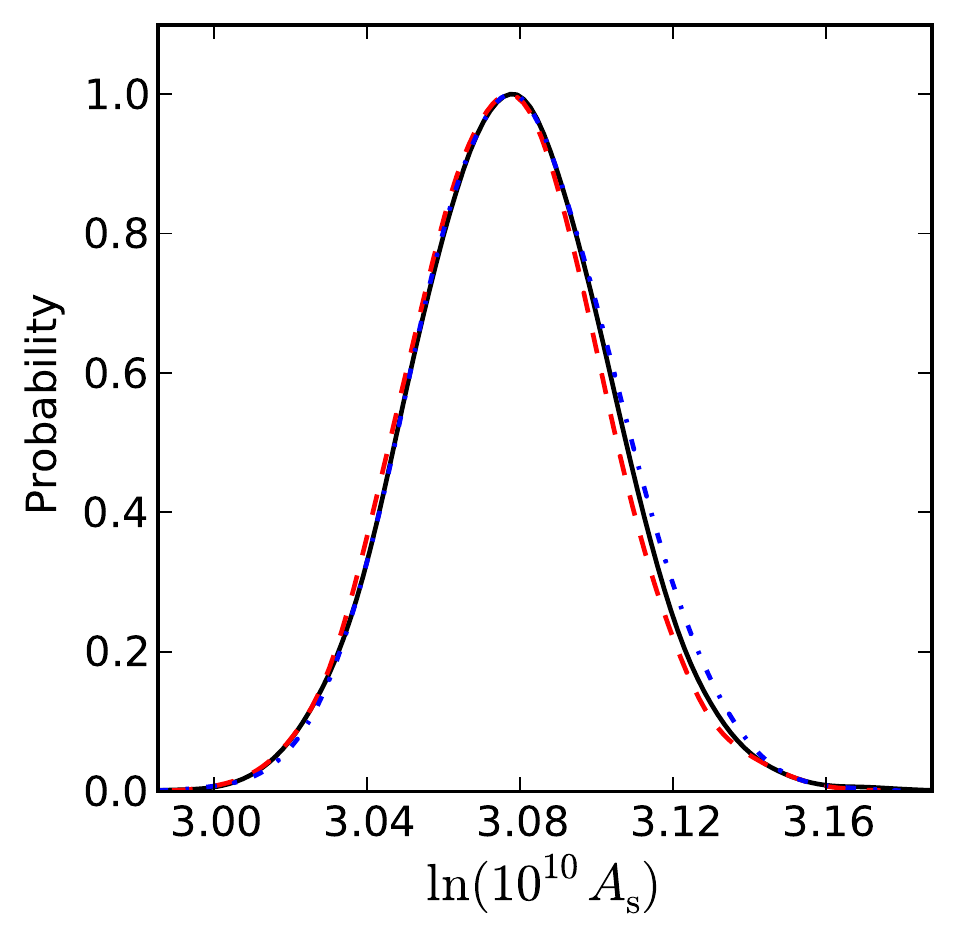}
%\end{figure*} 
%\begin{figure*}[htb!]
\includegraphics[width=3.0cm,]{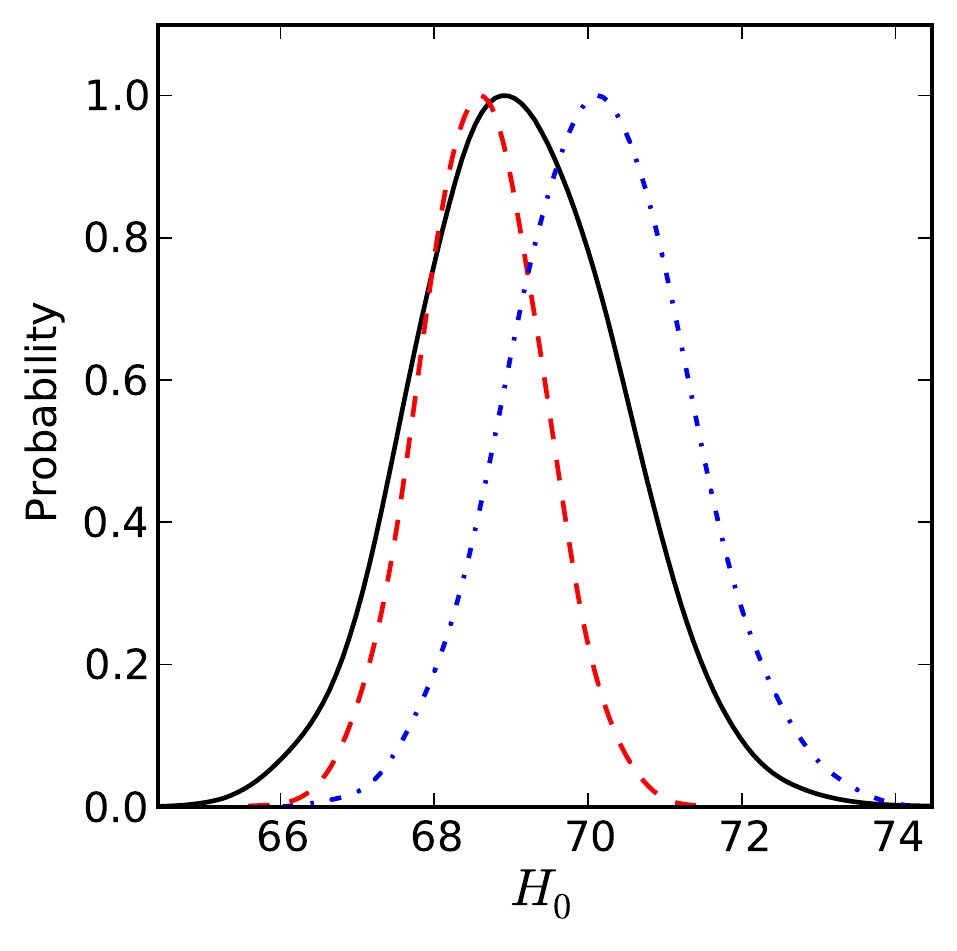}
\caption{\footnotesize Posterior distribution functions for the parameters described in the text when $A_{\rm L}=1$. 
We report PLANCK (solid line), PLANCK+BAO (dashed line) and PLANCK+HST (dashed-dotted line).}
\end{figure*} 

\begin{figure*}[htb!]
\includegraphics[width=3.0cm,]{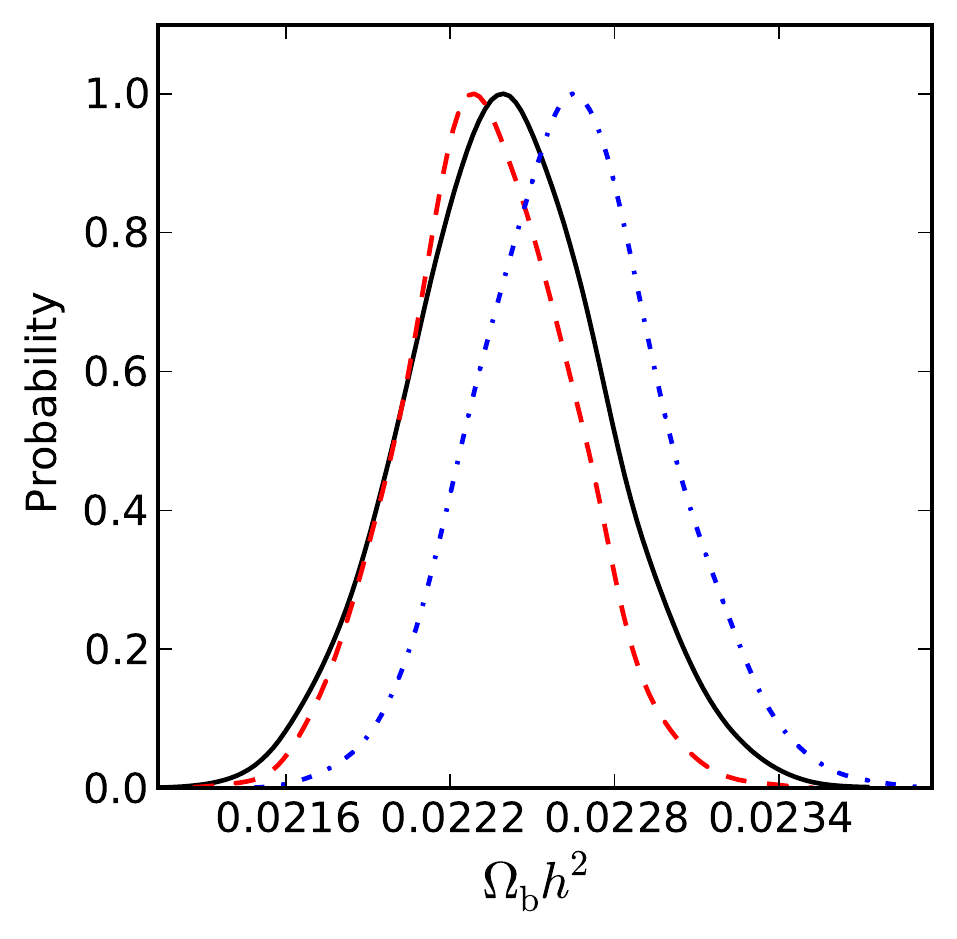}
\includegraphics[width=3.0cm,]{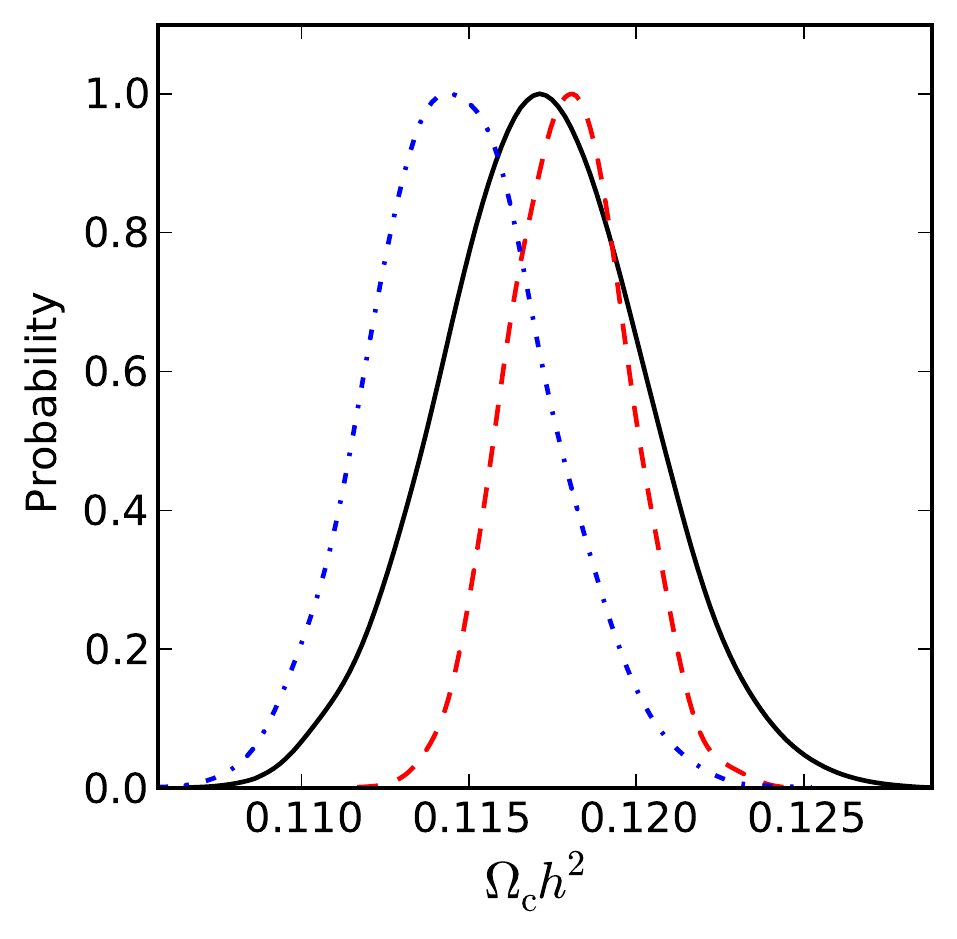}
\includegraphics[width=3.0cm,]{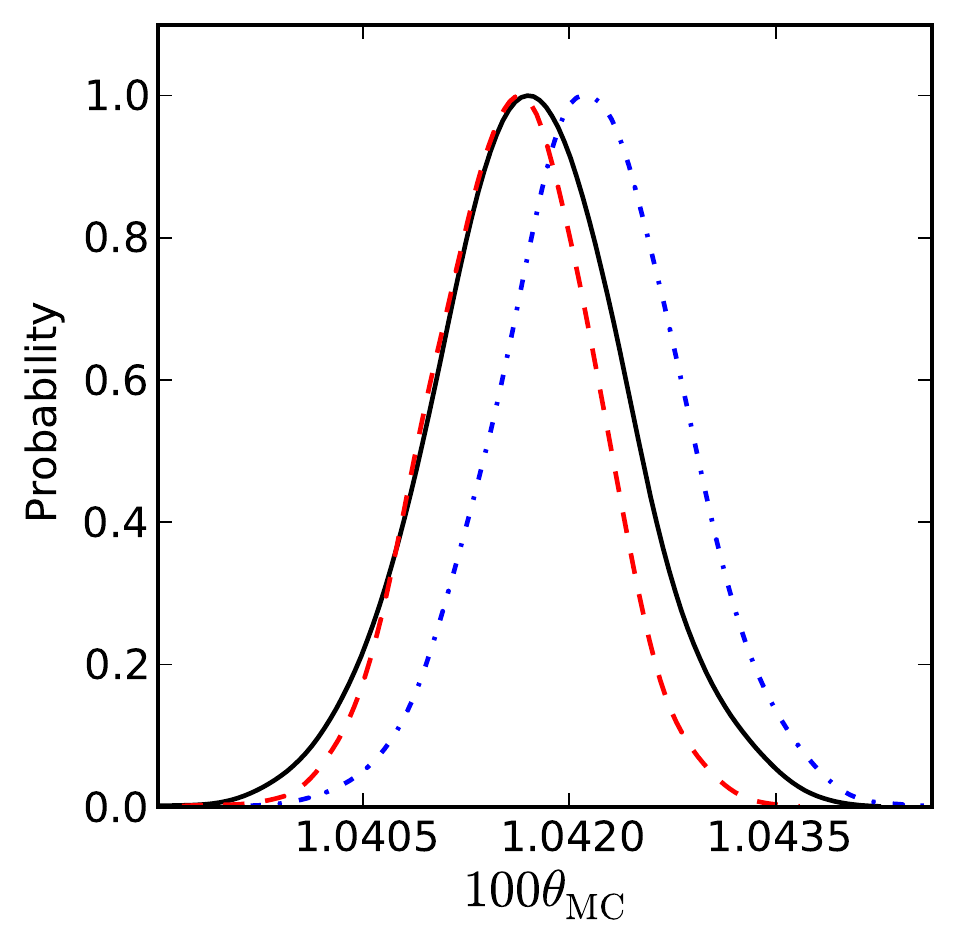}
%\end{figure*} 
%\begin{figure*}[htb!]
\includegraphics[width=3.0cm,]{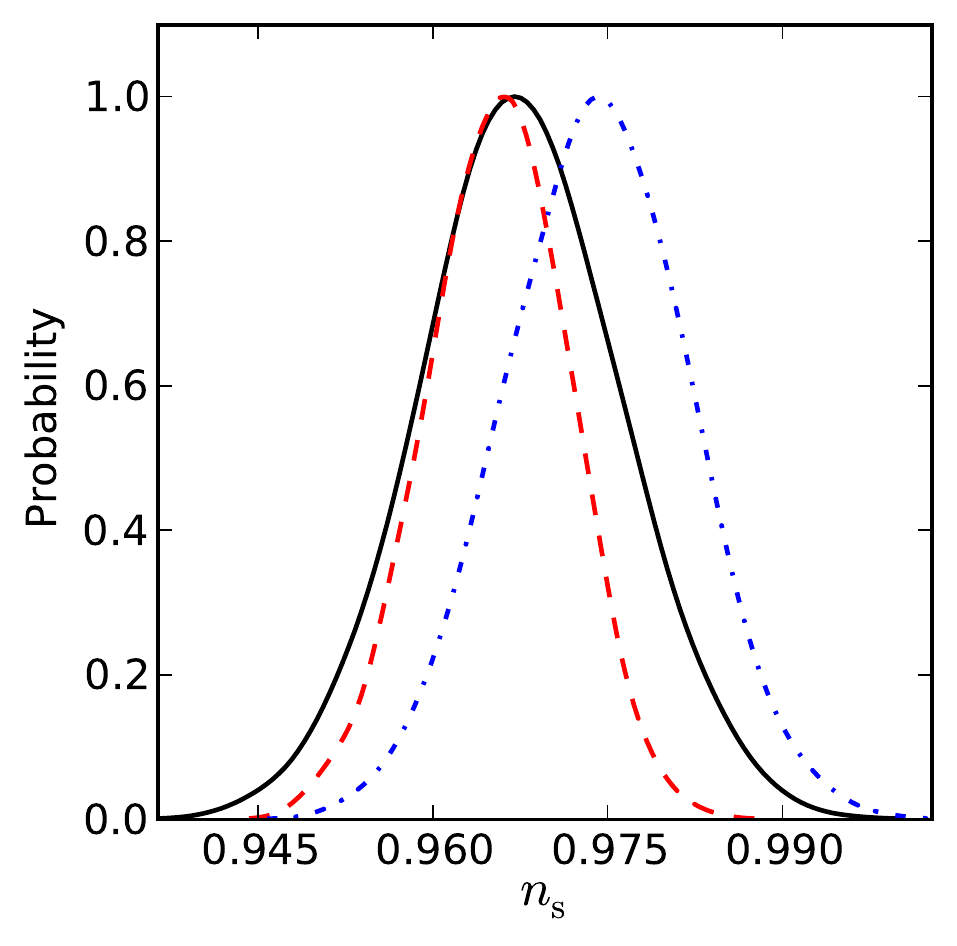}
\includegraphics[width=3.0cm,]{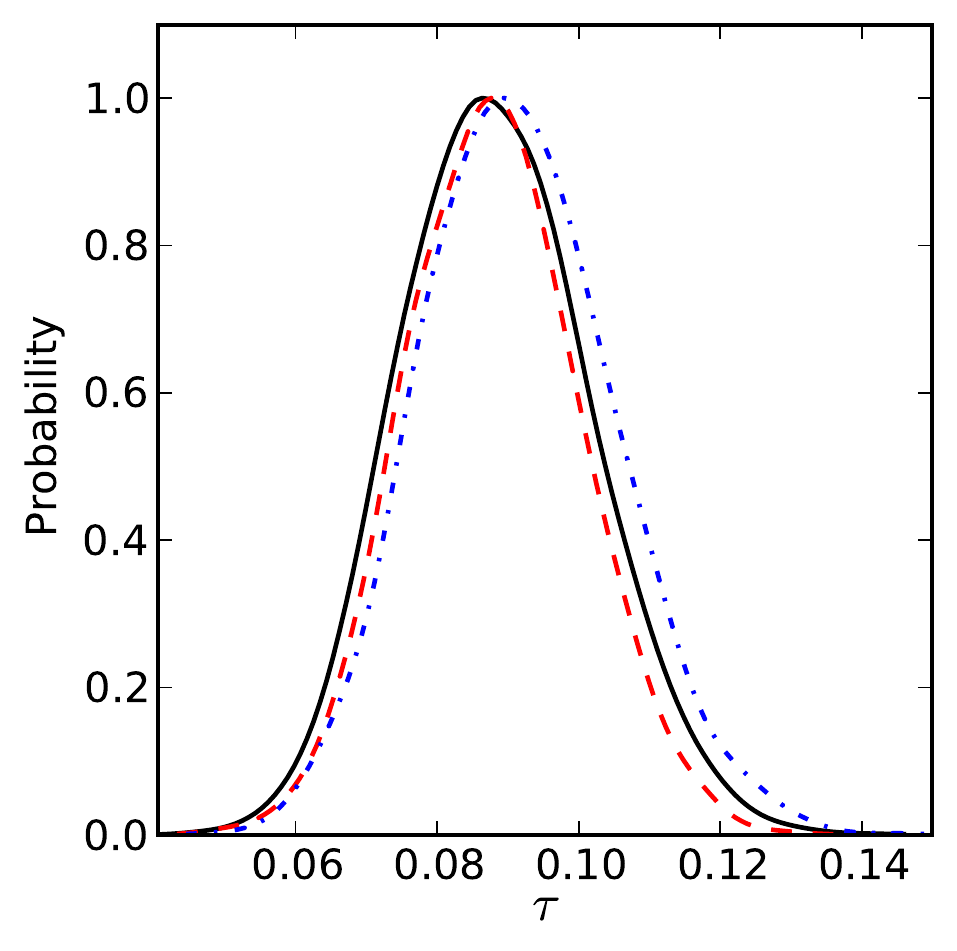}
\includegraphics[width=3.0cm,]{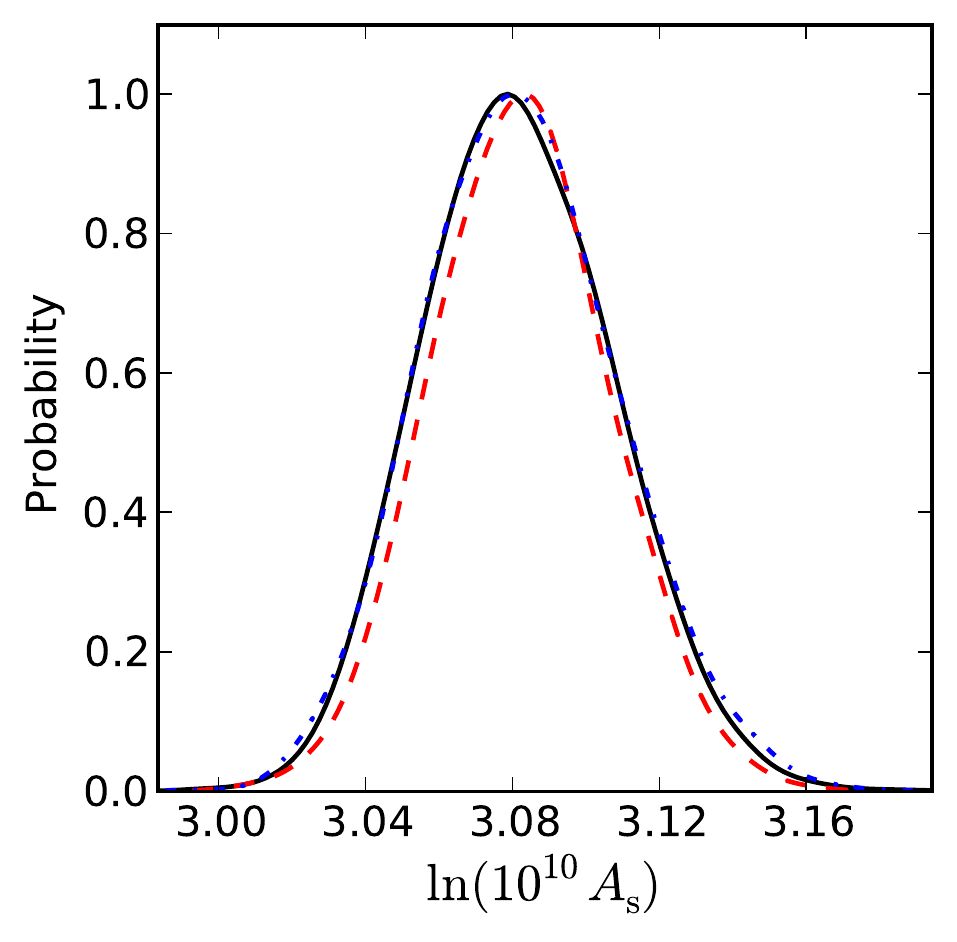}
%\end{figure*} 
%\begin{figure*}[htb!]
\includegraphics[width=3.0cm,]{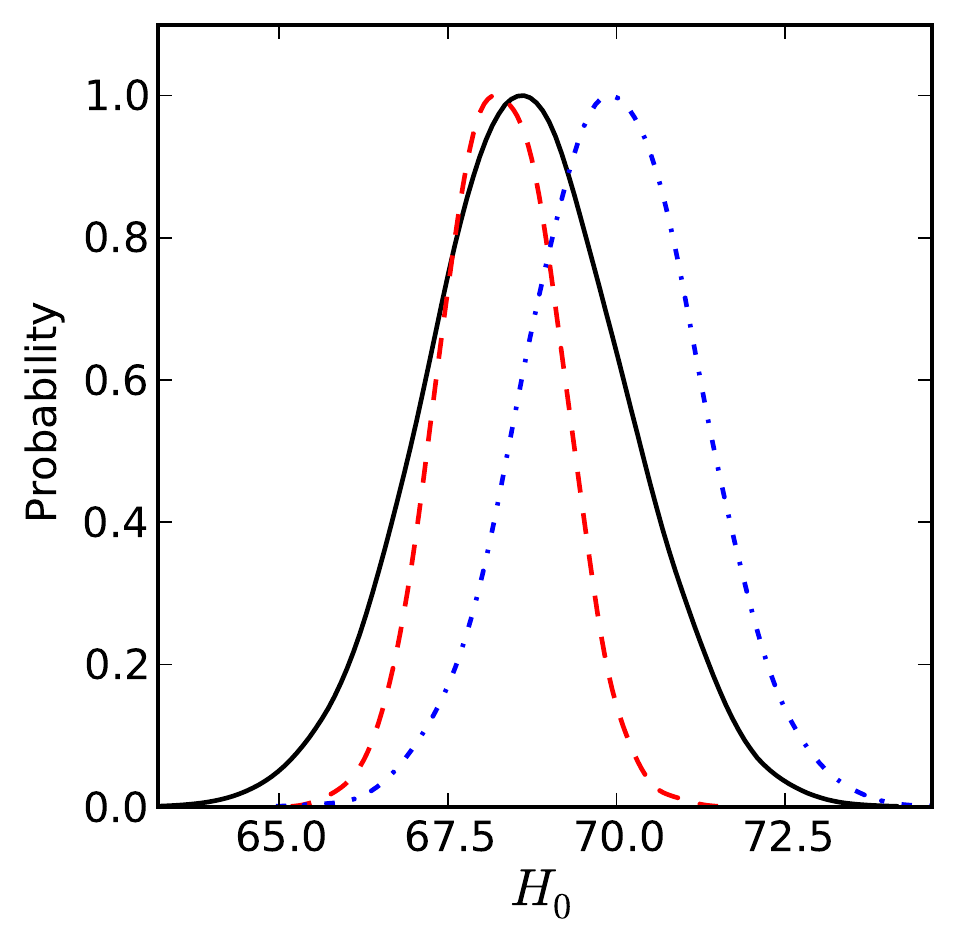}
\includegraphics[width=3.0cm,]{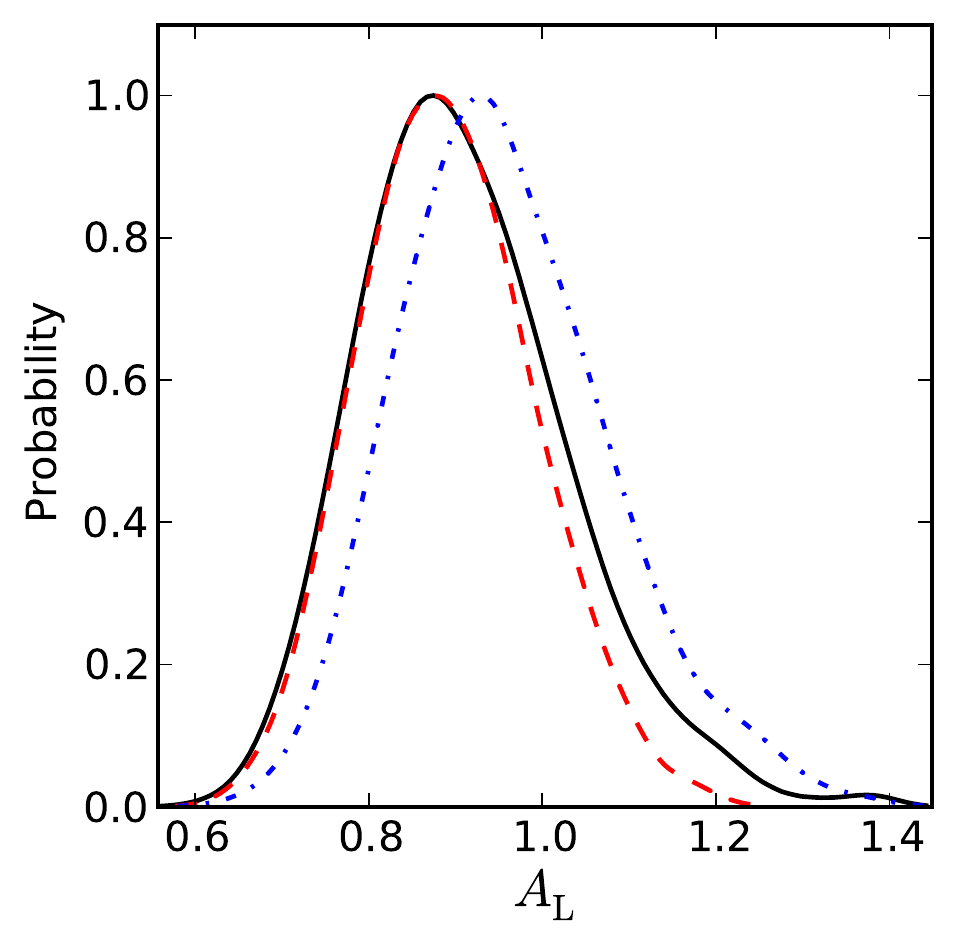}
\caption{\footnotesize Posterior distribution functions for the parameters described in the text when $A_{\rm}$ is free to vary. 
We report PLANCK (solid line), PLANCK+BAO (dashed line) and PLANCK+HST (dashed-dotted line).}
\end{figure*} 

\section {Conclusions} \label{sec:concl}
In this brief report we provide updated constraints on the length-scale parameter $B_{\rm 0}$ of \textit{f}$(R)$ theories, 
using the data recently released by the \PLANCK experiment together with the $H_{\rm 0}$ measurement from HST and BAO data sets
from four different surveys. We also investigate the effects of varying the lensing amplitude $A_{\rm L}$ from its
standard value. 

Our analysis provides the tightest constraint on $B_{\rm 0}$ from CMB measurements only and from one single experiment
in general ($B_{\rm 0}< 0.134$ 95\% c.l.). It also improves by a factor $\sim1.6$ the previous tighter constraint 
from the CMB measurements plus BAO data sets we reported in \cite{Marchini:2013lpp}. When we consider $A_{\rm L}=1$, the constraint
we obtain from PLANCK plus BAO is $B_{\rm 0}<0.085$ at 95\% c.l. Moreover we found a bimodal behavior for the $B_{\rm 0}$ posterior 
distribution when the HST prior is present or when $A_{\rm L}$ is free to vary, making the constraints weaker in these cases.

Furthermore we found that in the framework of the considered MG models the standard value of the lensing amplitude 
$A_{\rm L}=1$ returns to be in agreement with the \PLANCK measurements, oppositely to what happens in the $\Lambda$CDM scenario.

Another tantalizing feature we infer is the fact that MG scenario mitigates the \PLANCK-HST tension on the Hubble constant $H_{\rm 0}$ value.

%\clearpage

\end{document}